\documentclass{moriond}

\def\Journal#1#2#3#4{{#1} {\bf #2}, #3 (#4)}

\def\PLB{{\em Phys. Lett.}  B}
\def\PRL{\em Phys. Rev. Lett.}
\def\PRD{{\em Phys. Rev.} D}

\def\JHEP{\em JHEP}
\def\EPJC{{\em Eur. Phys. J.} C}

\def\be{\begin{equation}}
\def\ee{\end{equation}}
\def\bea{\begin{eqnarray}}
\def\eea{\end{eqnarray}}

\usepackage{amsmath}
\usepackage{ifthen} 

\newboolean{uprightparticles}
\setboolean{uprightparticles}{false} 

\usepackage{xspace} 
\usepackage{upgreek}

\def\lhcb   {\mbox{LHCb}\xspace}

\def\velo   {VELO\xspace}

\def\MagUp {\mbox{\em Mag\kern -0.05em Up}\xspace}

\ifthenelse{\boolean{uprightparticles}}%
{

 \def\Peta        {\ensuremath{\upeta}\xspace}

 \def\Ppi         {\ensuremath{\uppi}\xspace}

 \def\PDelta      {\ensuremath{\Delta}\xspace}                 
 \def\PXi         {\ensuremath{\Xi}\xspace}                 
 \def\PLambda     {\ensuremath{\Lambda}\xspace}                 
 \def\PSigma      {\ensuremath{\Sigma}\xspace}                 
 \def\POmega      {\ensuremath{\Omega}\xspace}                 
 \def\PUpsilon    {\ensuremath{\Upsilon}\xspace}

 \def\PB      {\ensuremath{\mathrm{B}}\xspace}                 
                  
 \def\PD      {\ensuremath{\mathrm{D}}\xspace}

 \def\PK      {\ensuremath{\mathrm{K}}\xspace}

 \def\Pb      {\ensuremath{\mathrm{b}}\xspace}

 \def\Pe      {\ensuremath{\mathrm{e}}\xspace}

 \def\Pi      {\ensuremath{\mathrm{i}}\xspace}

 \def\Ps      {\ensuremath{\mathrm{s}}\xspace}

 \def\thebaroffset{0.0em}
}
{

 \def\Peta        {\ensuremath{\eta}\xspace}

 \def\Ppi         {\ensuremath{\pi}\xspace}

 \mathchardef\PDelta="7101
 \mathchardef\PXi="7104
 \mathchardef\PLambda="7103
 \mathchardef\PSigma="7106
 \mathchardef\POmega="710A
 \mathchardef\PUpsilon="7107
                  
 \def\PB      {\ensuremath{B}\xspace}                 
                  
 \def\PD      {\ensuremath{D}\xspace}

 \def\PK      {\ensuremath{K}\xspace}

 \def\Pb      {\ensuremath{b}\xspace}

 \def\Pe      {\ensuremath{e}\xspace}

 \def\Pi      {\ensuremath{i}\xspace}

 \def\Ps      {\ensuremath{s}\xspace}

 \def\thebaroffset{0.18em}
}
\newcommand{\offsetoverline}[2][\thebaroffset]{\kern #1\overline{\kern -#1 #2}}%

\makeatletter
\ifcase \@ptsize \relax
  \newcommand{\miniscule}{\@setfontsize\miniscule{4}{5}}
\or
  \newcommand{\miniscule}{\@setfontsize\miniscule{5}{6}}
\or
  \newcommand{\miniscule}{\@setfontsize\miniscule{5}{6}}
\fi
\makeatother

\DeclareRobustCommand{\optbar}[1]{\shortstack{{\miniscule (\rule[.5ex]{1.25em}{.18mm})}
  \\ [-.7ex] $#1$}}

\def\epem       {{\ensuremath{\Pe^+\Pe^-}}\xspace}

\def\squark    {{\ensuremath{\Ps}}\xspace}

\def\bquark    {{\ensuremath{\Pb}}\xspace}

\def\pion   {{\ensuremath{\Ppi}}\xspace}
\def\piz    {{\ensuremath{\pion^0}}\xspace}
\def\pip    {{\ensuremath{\pion^+}}\xspace}
\def\pim    {{\ensuremath{\pion^-}}\xspace}

\def\kaon    {{\ensuremath{\PK}}\xspace}

\def\KorKbar {\kern \thebaroffset\optbar{\kern -\thebaroffset \PK}{}\xspace}

\def\Kp      {{\ensuremath{\kaon^+}}\xspace}
\def\Km      {{\ensuremath{\kaon^-}}\xspace}

\def\KS      {{\ensuremath{\kaon^0_{\mathrm{S}}}}\xspace}

\newcommand{\etaz}{\ensuremath{\Peta}\xspace}

\def\Dbar    {{\ensuremath{\offsetoverline{\PD}}}\xspace}
\def\D       {{\ensuremath{\PD}}\xspace}

\def\DorDbar {\kern \thebaroffset\optbar{\kern -\thebaroffset \PD}\xspace}
\def\Dz      {{\ensuremath{\D^0}}\xspace}
\def\Dzb     {{\ensuremath{\Dbar{}^0}}\xspace}
\def\Dp      {{\ensuremath{\D^+}}\xspace}
\def\Dm      {{\ensuremath{\D^-}}\xspace}

\def\DpDm    {\ensuremath{\Dp {\kern -0.16em \Dm}}\xspace}

\def\Dstarp  {{\ensuremath{\D^{*+}}}\xspace}

\def\Ds      {{\ensuremath{\D^+_\squark}}\xspace}
\def\Dsp     {{\ensuremath{\D^+_\squark}}\xspace}

\def\B       {{\ensuremath{\PB}}\xspace}

\def\BorBbar {\kern \thebaroffset\optbar{\kern -\thebaroffset \PB}\xspace}

\def\Bd      {{\ensuremath{\B^0}}\xspace}

\def\BdorBdbar {\kern \thebaroffset\optbar{\kern -\thebaroffset \Bd}\xspace}

\def\Bs      {{\ensuremath{\B^0_\squark}}\xspace}

\def\BsorBsbar {\kern \thebaroffset\optbar{\kern -\thebaroffset \Bs}\xspace}

\def\Y#1S{\ensuremath{\PUpsilon{(#1S)}}\xspace}

\def\LorLbar     {\kern \thebaroffset\optbar{\kern -\thebaroffset \PLambda}\xspace}

\newcommand{\decay}[2]{\ensuremath{#1\!\to #2}\xspace} 

\def\to                 {\ensuremath{\rightarrow}\xspace}

\def\CP                {{\ensuremath{C\!P}}\xspace}

\def\AT#1     {\ensuremath{A_{\mathrm{T}}^{#1}}\xspace}           

\def\C#1      {\ensuremath{\mathcal{C}_{#1}}\xspace}                       
\def\Cp#1     {\ensuremath{\mathcal{C}_{#1}^{'}}\xspace}                    
\def\Ceff#1   {\ensuremath{\mathcal{C}_{#1}^{\mathrm{(eff)}}}\xspace}        
\def\Cpeff#1  {\ensuremath{\mathcal{C}_{#1}^{'\mathrm{(eff)}}}\xspace}       
\def\Ope#1    {\ensuremath{\mathcal{O}_{#1}}\xspace}                       
\def\Opep#1   {\ensuremath{\mathcal{O}_{#1}^{'}}\xspace}                    

\newcommand{\aunit}[1]{\ensuremath{\text{\,#1}}}       

\newcommand{\tev}{\aunit{Te\kern -0.1em V}\xspace}
\newcommand{\gev}{\aunit{Ge\kern -0.1em V}\xspace}
\newcommand{\mev}{\aunit{Me\kern -0.1em V}\xspace}
\newcommand{\kev}{\aunit{ke\kern -0.1em V}\xspace}
\newcommand{\ev}{\aunit{e\kern -0.1em V}\xspace}
 
\newcommand{\mevc}{\ensuremath{\aunit{Me\kern -0.1em V\!/}c}\xspace}
\newcommand{\gevc}{\ensuremath{\aunit{Ge\kern -0.1em V\!/}c}\xspace}
\newcommand{\mevcc}{\ensuremath{\aunit{Me\kern -0.1em V\!/}c^2}\xspace}
\newcommand{\gevcc}{\ensuremath{\aunit{Ge\kern -0.1em V\!/}c^2}\xspace}

\def\gsim{{~\raise.15em\hbox{$>$}\kern-.85em
          \lower.35em\hbox{$\sim$}~}\xspace}
\def\lsim{{~\raise.15em\hbox{$<$}\kern-.85em
          \lower.35em\hbox{$\sim$}~}\xspace}

\def\tell1  {TELL1\xspace}
\def\ukl1   {UKL1\xspace}

\newcommand{\Photo}{}

\begin{document}
\vspace*{4cm}
\title{Mixing and \CP violation in charm decays at \lhcb}

\author{Tom Hadavizadeh, on behalf of the \lhcb collaboration}

\address{Monash University, Melbourne, Australia }


\maketitle\abstracts{
Recent measurements of mixing and \CP violation in charm decays at \lhcb are presented. These include searches for direct \CP violation in $\decay{\Dz}{\KS\KS}$, $\decay{D_{(s)}^{+}}{h^{+}\piz}$ and $\decay{D_{(s)}^{+}}{h^{+}\etaz}$ decays, and a search for time-dependent \CP violation in \decay{\Dz}{h^{+}h^{-}} decays, where $h^{+}$ is either a \pip or \Kp meson. 
}

\section{Introduction}
Charm hadrons provide a unique opportunity to test \CP violation in the decays of up-type quarks. 
However, due to the small size of the relevant CKM matrix elements, \CP violation in charm decays is very suppressed and typically of the order $10^{-4}$--$10^{-3}$. 
For both neutral and charged \D mesons, \CP violation can manifest itself directly as an asymmetry in the decay rates of particle and antiparticle states. For neutral \D mesons, the oscillation between \Dz and \Dzb flavour eigenstates enables indirect \CP violation in the mixing, and in the interference between mixing and decay. 

Large samples of \D mesons have been collected by the \lhcb experiment during Run 1 and Run 2 of the LHC. 
Direct \CP violation has been observed in neutral \Dz mesons in the difference of the time-integrated \CP asymmetries $\Delta \mathcal{A}_{\CP} = \mathcal{A}_{\CP}(\decay{\Dz}{\Kp\Km}) - \mathcal{A}_{\CP}(\decay{\Dz}{\pip\pim})$.\cite{LHCb-PAPER-2019-006}
However, further measurements are required to help shed light on the nature of this observation, which, to be explained within the Standard Model (SM), requires an enhancement of rescattering effects by one order of magnitude beyond the naive QCD expectation. 
Three recent such measurements are presented. 

\section{Measurement of \texorpdfstring{\CP}{CP} asymmetry in \texorpdfstring{$\decay{\Dz}{\KS\KS}$}{D02KS0KS0} decays}

The decay \decay{\Dz}{\KS\KS} is an ideal candidate for observing \CP violation in charm decays as it only receives contributions from similarly sized loop-suppressed and tree-level exchange diagrams that disappear in the flavour-$\text{SU}(3)$ limit.
Theory predictions estimate that $\mathcal{A}_{\CP}(\decay{\Dz}{\KS\KS})$ could be as large as $\mathcal{O}(\%)$.\cite{Nierste:2015zra}
The results presented here use the full Run 2 data set,\cite{LHCb-PAPER-2020-047} including a reanalysis of the data taken during 2015--2016 with 30\% improved sensitivity over the previous measurement.\cite{LHCb-PAPER-2018-012} The \Dz mesons originate from $\decay{\Dstarp}{\Dz\pip}$ decays, allowing the flavour of the \Dz to be determined from the charge of the accompanying pion. 
Only a fraction of \KS mesons decay early enough to be within the \velo tracking detector, but they have a better mass, momentum and vertex resolution. 
Each \KS meson is reconstructed using two long (L) or downstream (D) tracks, which include or do not include segments in the \velo tracking detector. This leads to three categories when the \KS mesons are combined to form \Dz mesons, which are referred to as LL, LD and DD. 
The time-integrated $\mathcal{A}_{\CP}$ is measured using a simultaneous maximum-likelihood fit to \mbox{$\Delta m = m(\KS\KS\pip) - m(\KS\KS)$} and the masses of both \KS candidates.

\begin{figure}[tb]
    \centering
    \includegraphics[width=0.32\linewidth]{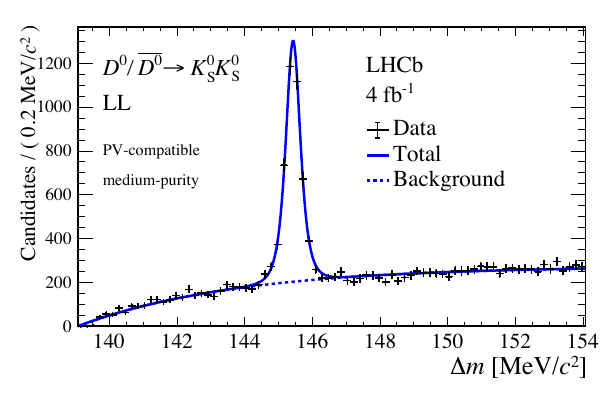}
    \includegraphics[width=0.32\linewidth]{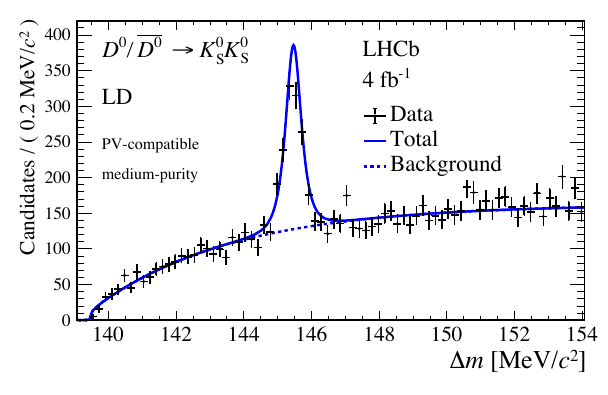}
    \includegraphics[width=0.32\linewidth]{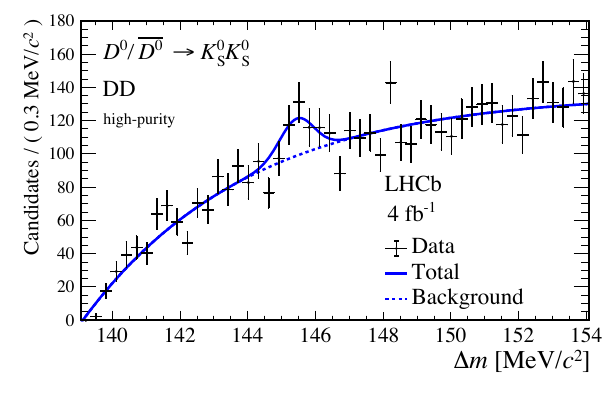}
        \vspace{-0.3cm}
    \caption{Distribution of the $\Delta m$ observable for $\decay{\Dz}{\KS\KS}$ candidates in the (left) LL, (middle)  LD and (right) DD \KS reconstruction categories. Fit projections are overlaid. }
    \label{fig:D02KSKS_fits}
\end{figure}

This measurement includes improvements in the analysis strategy that increase the sensitivity with respect to previous measurements. 
The candidates are split according to whether they are compatible with originating at the primary proton--proton interaction vertex (PV); for those that are, a factor of two better mass resolution is achieved by constraining the \Dstarp decay vertex to coincide with the PV. 
A multivariate classifier (kNN) is used to reduce the combinatorial background, and additionally to split the remaining candidates into categories, improving the sensitivity to $\mathcal{A}_{\CP}$.
The effect of the production and detection nuisance asymmetries are removed using input from a calibration sample of $\decay{\Dz}{\Kp\Km}$ decays, accounting for the different kinematics through a momentum-dependent weighting. 
Examples of the distribution of $\Delta m$ in the different \KS reconstruction categories are shown in Fig.~\ref{fig:D02KSKS_fits}.

The \CP asymmetry is measured to be 
 \begin{equation*}
    \mathcal{A}_{\CP}(\decay{\Dz}{\KS\KS}) = (-3.1\pm 1.2 \pm 0.4 \pm 0.2)\%,
\end{equation*}
where the first uncertainty is statistical, the second systematic and the third arises from the uncertainty on $\mathcal{A}_{\CP}(\decay{\Dz}{\Kp\Km})$.\cite{LHCb-PAPER-2016-035} 
This constitutes the most precise measurement to date, is compatible with no \CP asymmetry within 2.4 standard deviations and is in agreement with previous determinations.

\section{Search for \texorpdfstring{\CP}{CP} violation in \texorpdfstring{$\decay{D_{(s)}^{+}}{h^{+}\piz}$}{D2HPi0} and \texorpdfstring{$\decay{D_{(s)}^{+}}{h^{+}\etaz}$}{D2HEta} decays}
The two-body decays of charged \D mesons $\decay{D_{(s)}^{+}}{h^{+}\piz}$ and $\decay{D_{(s)}^{+}}{h^{+}\etaz}$, where $h^{+}$ stands for \pip or \Kp, proceed via Cabibbo-favoured, singly Cabibbo-suppressed (SCS) or doubly Cabibbo-suppressed amplitudes. The SCS decays $\decay{\Ds}{\Kp}{\piz}$, $\decay{\D}{\pip}{\etaz}$ and $\decay{\Ds}{\Kp}{\etaz}$ are expected to have a \CP asymmetry $\mathcal{O}(10^{-4}\textnormal{--}10^{-3})$.\cite{PhysRevD.85.034036,PhysRevD.86.036012,Pirtskhalava:2011va,PhysRevD.99.113001}
Due to isospin constraints, the SCS mode \mbox{$\decay{\Dp}{\pip\piz}$} is expected to have a \CP asymmetry  smaller than $10^{-5}$ in the SM; a nonzero \CP asymmetry would indicate physics beyond the SM.\cite{Grossman:2012eb}

\begin{figure}[tb]
    \centering
    \includegraphics[width=0.4\linewidth]{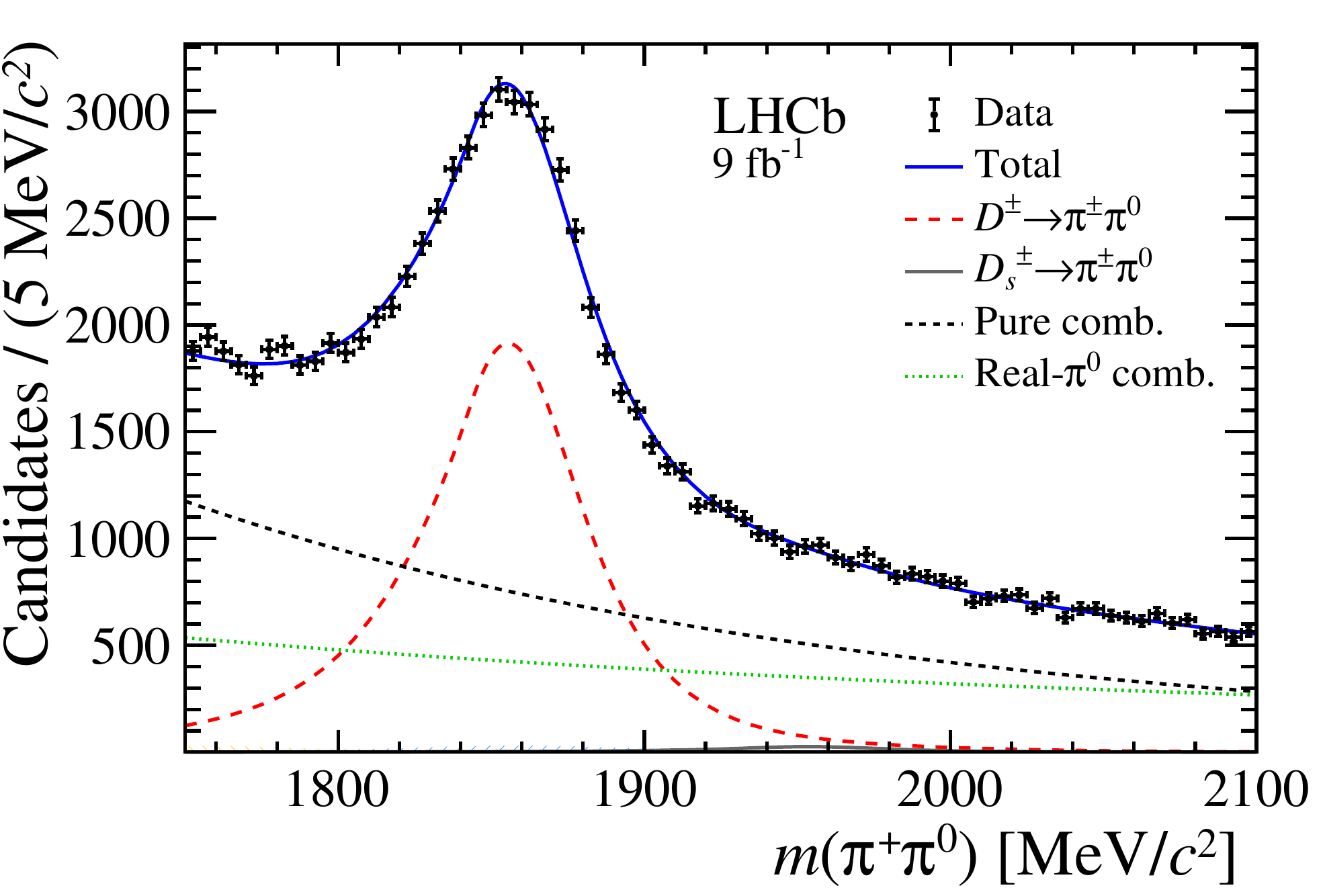}
    \includegraphics[width=0.4\linewidth]{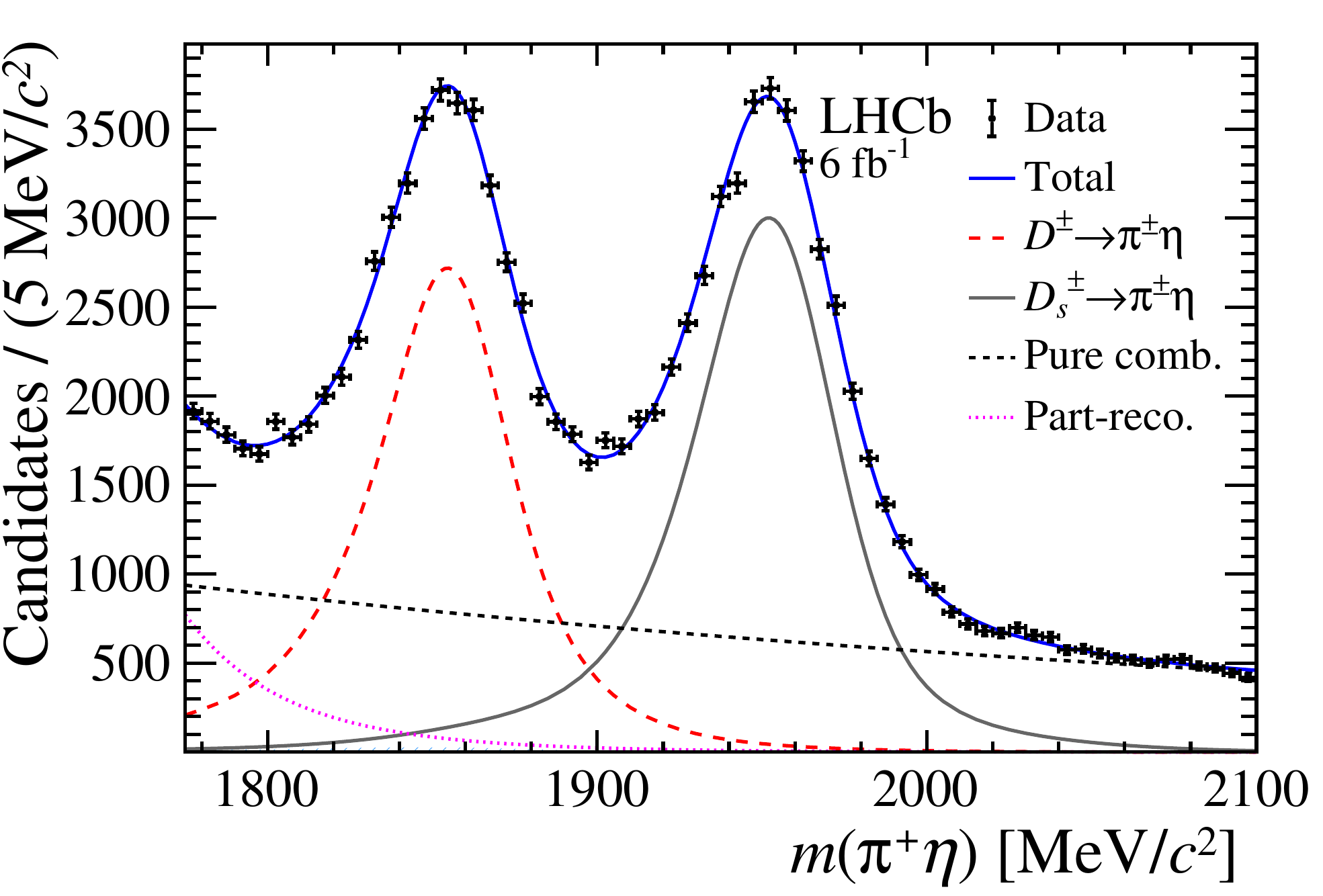}
    \vspace{-0.3cm}
    \caption{Distribution of $m(h^+h^0)$ for (left) $\decay{D_{(s)}^{+}}{\pip\piz}$ and (right) $\decay{D_{(s)}^{+}}{\pip\etaz}$ candidates. Fit projections are overlaid.}
    \label{fig:D2hh0_fits}
\end{figure}

Modes with only one track and a neutral particle are challenging to reconstruct at \lhcb as it is not possible to determine the displaced \D-meson decay vertex to suppress background.
Reconstructing decays of the neutral mesons into the $\epem\gamma$ final state enables the \D decay vertex to be determined; however, the branching fractions for these processes are suppressed by two orders of magnitude with respect to the two-photon final state. A contribution, larger by a factor of four, to the same final state arises from decays of the neutral mesons to two photons, where one photon interacts with the detector material and converts into an $\epem$ pair.
The \CP asymmetry measurements are performed using data taken during Run 1 and (or) Run 2 of the LHC for final states including a \piz (\etaz) meson, using two-dimensional fits to the invariant masses $m(\epem\gamma)$ and $m(h^+h^0) \equiv m(h^{+}\epem\gamma) - m(\epem\gamma) + m(h^{0}) $, where $m(h^{0})$ is the known neutral-meson mass.\cite{LHCb-PAPER-2021-001} The $m(h^+h^0)$ distributions are shown in Fig~\ref{fig:D2hh0_fits}.
The production and detection nuisance asymmetries are subtracted using large samples of $\decay{D_{(s)}^{+}}{\KS h^+}$ decays, for which the \CP asymmetry is know with a better precision.\cite{LHCb-PAPER-2019-002}
The $\decay{D_{(s)}^{+}}{\KS h^+}$ samples are weighted to match the kinematics of the signal samples to ensure cancellation of the nuisance asymmetries.  
The results are consistent with no \CP violation in all studied modes: 

\noindent\begin{minipage}{.5\linewidth}
\begin{alignat*}{7}
    \mathcal{A}_{\CP}(\decay{\Dp}{\pip\piz}) 	&= (-&&1.3 &&\pm 0.9 &&\pm 0.6 &)\%, \\
    \mathcal{A}_{\CP}(\decay{\Dp}{\Kp\piz}) 	&= (-&&3.2 &&\pm 4.7 &&\pm 2.1 &)\%, \\
    \mathcal{A}_{\CP}(\decay{\Dp}{\pip\etaz})   &= (-&&0.2 &&\pm 0.8 &&\pm 0.4 &)\%, \\
    \mathcal{A}_{\CP}(\decay{\Dp}{\Kp\etaz}) 	&= (-&&6 &&\pm 10 &&\pm 4 &)\%.\\
\end{alignat*}
\end{minipage}%
\begin{minipage}{.5\linewidth}
\begin{alignat*}{7}
    \mathcal{A}_{\CP}(\decay{\Dsp}{\Kp\piz}) 	&= (-&&0.8 &&\pm 3.9 &&\pm 1.2 &)\%, \\
    \mathcal{A}_{\CP}(\decay{\Dsp}{\pip\etaz})  &= (&&0.8 &&\pm 0.7 &&\pm 0.5 &)\%, \\
    \mathcal{A}_{\CP}(\decay{\Dsp}{\Kp\etaz})   &= (&&0.9 &&\pm 3.7 &&\pm 1.1 &)\%,\\
    \\
\end{alignat*}
\end{minipage}
Five of these constitute the most precise measurements to date.

\section{Search for time-dependent \texorpdfstring{\CP}{CP} violation in \texorpdfstring{\decay{\Dz}{\Kp\Km}}{D02KK} and \texorpdfstring{\decay{\Dz}{\pip\pim}}{D02PiPi} decays}

The \CP asymmetry of neutral \D mesons can vary as a function of decay time as a result of the mixing between \Dz and \Dzb mesons. The mixing can be quantified using the parameters $x_{12}\equiv 2|M_{12}/\Gamma|$ and $y_{12}\equiv|\Gamma_{12}/\Gamma|$,\cite{Grossman:2009mn} where $\mathbf{M}- \frac{i}{2}\mathbf{\Gamma}$ is the effective Hamiltonian of the \Dz-meson system and $\Gamma$ is the average decay width of the mass eigenstates $D_{1,2}$.  
The asymmetry of the time-dependent decay rates into the final state $f=\Kp\Km$ or $\pip\pim$ can be approximated as
\begin{equation}
    \mathcal{A}_{\CP}(\decay{\Dz}{f},t)\approx a^{d}_{f} + \Delta Y_{f} \frac{t}{\tau_{\Dz}},
\end{equation}
where $a^{d}_{f}$ is the \CP violation in decay, $\tau_{\Dz}$ is the \Dz meson lifetime, and the slope $\Delta Y_{f}$ is approximately equal to the negative of the parameter $A_{\Gamma}^{f}$ used in previous measurements.\cite{Staric:2015sta}
The quantity $\Delta Y_{f}$ is expected to be in the range $10^{-4}\textnormal{--}10^{-5}$ in the SM.\cite{Kagan:2020vri,Li:2020xrz}
The current world average \mbox{$ \Delta Y =  (3.1\pm2.0)\times10^{-4}$}, which neglects subleading final-state dependent contributions, shows no evidence of \CP violation in the mixing.\cite{Amhis:2019ckw}

The new measurements presented here use the full Run 2 data set.\cite{LHCb-PAPER-2020-045}
The analysis method is developed and validated using the decay \decay{\Dz}{\Km\pip}, for which the analogue of $\Delta Y$ is known to be smaller than the current experimental precision. 
The data sample comprises 58 million $\decay{\Dz}{\Kp\Km}$ and 18 million $\decay{\Dz}{\pip\pim}$ signal candidates and 0.5 billion \decay{\Dz}{\Km\pip} control candidates, achieving a precision of $0.5\times10^{-4}$ in the parameter $\Delta Y_{\Km\pip}$. The \Dz mesons are required to originate from the $\decay{\Dstarp}{\Dz\pip}$ decay.
The selection requirements induce correlations between the momenta and decay time, causing the momentum-dependent detection asymmetries to result in a nonlinear raw asymmetry as a function of decay time. 
Equalising the kinematic distributions of \Dz and \Dzb candidates and of \pip and \pim candidates successfully removes the time dependence in the control-mode asymmetry. 
Contributions from \Dz mesons from \bquark-hadron decays are removed by placing a requirement on the impact parameter of the \Dz meson. A small fraction remains (4\%), which increases as a function of decay time.
The associated bias on $\Delta Y$, \mbox{$0.3\times10^{-4}$}, is measured and subtracted.

\begin{figure}[tb]
    \centering
    \includegraphics[width=0.49\linewidth]{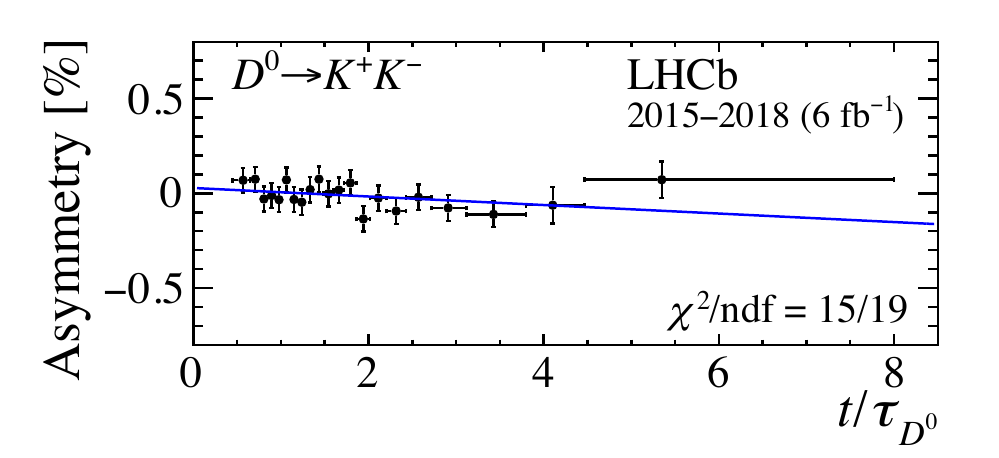}
    \includegraphics[width=0.49\linewidth]{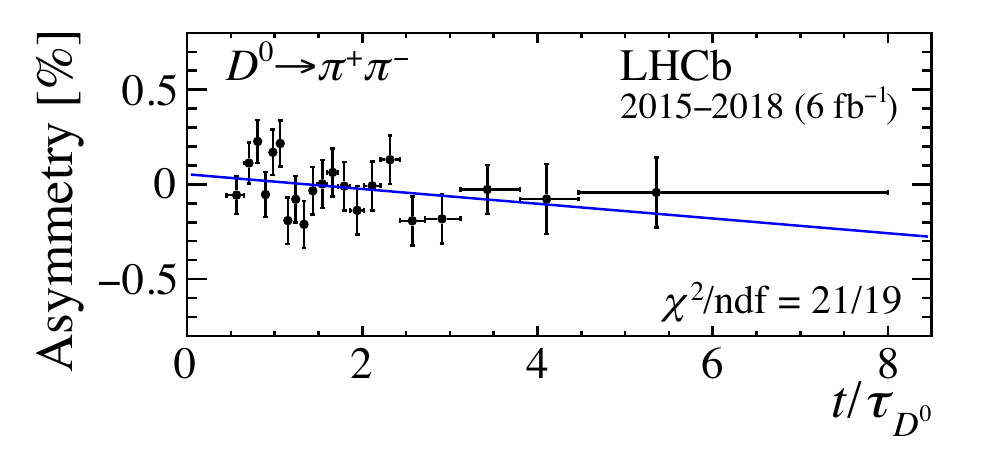}
    \vspace{-0.4cm}
    \caption{Linear fit to the time-dependent asymmetry of the (left) $\decay{\Dz}{\Kp\Km}$ and (right) $\decay{\Dz}{\pip\pim}$ candidates. }
    \label{fig:D02hh_time}
\end{figure}

The fits to the asymmetry of the signal candidates as a function of decay time are shown in Fig.~\ref{fig:D02hh_time}. 
The results are
\begin{align*}
    \Delta Y_{\Kp\Km}   &= (-2.3\pm 1.5\pm 0.3) \times 10^{-4},\\
    \Delta Y_{\pip\pim} &= (-4.0\pm 2.8\pm 0.4) \times 10^{-4},
\end{align*}
which are combined with  previous \lhcb measurements to yield the \lhcb legacy result with Run~1 and Run~2 data, \mbox{$\Delta Y = (-1.0\pm1.1\pm0.3)\times10^{-4}$}. This is consistent with no \CP violation and improves on the precision of the world average by a factor of nearly two. 

\section{Conclusions}

Searches for direct \CP asymmetry have been performed in $\decay{\Dz}{\KS\KS}$, $\decay{D_{(s)}^{+}}{h^{+}\piz}$ and $\decay{D_{(s)}^{+}}{h^{+}\etaz}$ decays, leading to worlds best measurements in most modes. All of the results are consistent with no \CP violation. Time-dependent \CP violation parameters are measured in $\decay{\Dz}{h^{+}h^{-}}$ decays and found to be consistent with no \CP violation, greatly improving the precision with respect to the current world average.
Many more charm measurements are underway with the Run 2 dataset and the upgraded \lhcb detector in Run 3 is expected to collect even larger samples.

\section*{References}


\end{document}